\documentclass[12pt]{webofc}
\usepackage[T1]{fontenc}
\usepackage[latin9]{inputenc}
\usepackage{amsmath}
\usepackage{graphicx}

\makeatletter
\usepackage[varg]{txfonts}
\makeatother

\begin{document}
\title{Studying particle acceleration from driven magnetic reconnection at the termination shock of a relativistic striped wind using particle-in-cell simulations}
\author{
\firstname{Yingchao} \lastname{Lu}
\inst{1,2}
\fnsep\thanks{\email{yclu@lanl.gov}} 
\and         
\firstname{Fan} \lastname{Guo}
\inst{1}
\and         
\firstname{Patrick} \lastname{Kilian}
\inst{1}
\and
\firstname{Hui} \lastname{Li}
\inst{1}
\and
\firstname{Chengkun} \lastname{Huang}
\inst{1}
\and
\firstname{Edison} \lastname{Liang}
\inst{2}
}

\institute{
Theoretical Division, Los Alamos National Laboratory, Los Alamos, New Mexico, 87545, USA
\and
Department of Physics and Astronomy, Rice University, Houston, Texas 77005, USA
}

\abstract{A rotating pulsar creates a surrounding pulsar wind nebula (PWN) by steadily releasing an energetic wind into the interior of the expanding shockwave of supernova remnant or interstellar medium. At the termination shock of a PWN, the Poynting-flux-dominated relativistic striped wind is compressed. Magnetic reconnection is driven by the compression and converts magnetic energy into particle kinetic energy and accelerating particles to high energies. We carrying out particle-in-cell (PIC) simulations to study the shock structure as well as the energy conversion and particle acceleration mechanism. By analyzing particle trajectories, we find that many particles are accelerated by Fermi-type mechanism. The maximum energy for electrons and positrons can reach hundreds of TeV.}

\maketitle

\section{Introduction}

In recent observations, high-energy emissions have been detected in
both young pulsar wind nebulae (PWNe) such as the Crab nebula and
mid-age PWNe such as the Geminga nebula\cite{Positron_CR_Yueksel2009}.
The high-energy emissions is likely due to the high-energy electrons
scattering off the cosmic microwave background (CMB) photons\cite{VHE_Amenomori2019,VHE_Abeysekara2019}.
The high-energy electrons and positrons need to be produced by a particle
acceleration in the nebula\cite{Positron_Abeysekara2017}. How pulsar
winds efficiently accelerate electrons and positrons to high energies
is a major puzzle and holds the key of understanding the near-earth
positron anomaly\cite{Positron_CR_Yueksel2009,Positron_Accardo2014,Positron_Hooper2017,Positron_Abeysekara2017}
as well as gamma rays from the Galactic Center\cite{CR_Abdo2007,Positron_CR_Yueksel2009,CR_Linden2018}.
In the synchrotron spectra of PWNe, the spectral break between the
radio and the X-ray cannot be explained by synchrotron cooling and
is believed to be attributed to the particle acceleration at the termination
shock\cite{TSacc_Rees1974,TSacc_Kennel1984,TSacc_MHD_Kennel1984}.
How the electromagnetic energy is converted into non-thermal particle
acceleration at the termination shock is not fully understood.

In the case of an obliquely rotating pulsar, a radially propagating
relativistic flow is continuously launched. Asymptoticly near the
equatorial plane, such a flow is modeled as a striped wind containing
a series of drifting Harries current sheets\cite{Striped_Coroniti1990,Striped_Kirk2003}
with opposite polarity magnetic fields in between. Numerical simulations
including magnetohydrodynamics\cite{TSacc_MHD_Kennel1984,MHD_Porth2014,MHD_Olmi2015,MHD_Porth2016},
particle-in-cell (PIC)\cite{Sironi2011} and test-particle simulations\cite{Trajectory_Giacinti2018}
have been used for modeling the termination shock of PWNe. Magnetic
reconnection driven by the termination shock may dissipate the magnetic
energy and accelerates particles\cite{Striped_Petri2007,Sironi2011}.
Particle acceleration in relativistic magnetic reconnection has been
a recent topic of strong interests. Two candidate mechanisms of particle
acceleration\cite{Reconn_Sironi2014,Reconn_Guo2014,Reconn_Guo2015,Reconn_Werner2015,Reconn_Guo2019}
are direct acceleration surrounding X-points and Fermi acceleration
in flows generated within the reconnection layer. While several analyses\cite{Reconn_Sironi2014,Reconn_Guo2014,Reconn_Guo2015,Reconn_Werner2015}
have shown that Fermi acceleration dominates particle acceleration
to high energies in a spontaneous reconnection with weak guide field,
this has not been studied in a shock-driven reconnection.

In this paper, we carry out two-dimensional particle-in-cell (PIC)
simulations to model the relativistic striped wind interacting with
the termination shock near the equatorial plane of obliquely rotating
pulsars. The magnetic reconnection driven by the precursor perturbation
from the shock converts the magnetic energy into particle energy and
accelerates particles into a power-law energy spectrum. The maximum
energy for electrons and positrons can reach hundreds of TeV if the
wind has a bulk Lorentz factor $\gamma_{0}\approx10^{6}$ and upstream
magnetization parameter $\sigma_{0}=10$.

\section{Simulation setup}

We use the PIC code EPOCH2D\cite{EPOCH_Arber2015} to study the kinetic
processes in the termination shock of a relativistic striped wind.
To ensure that the magnetic reconnection is driven by the physical
perturbation from the shock instead of numerical instabilities such
as numerical Cherenkov instability (NCI), the code is modified by
the authors to use a piecewise polynomial force interpolation scheme
with time-step dependency\cite{NCI_Lu2019}. The relativistic striped
wind is a steady magnetized electron-positron flow propagating along
$-\hat{\boldsymbol{x}}$. The spatial profile of the electromagnetic
field in the downstream rest frame $S$ is
\begin{equation}
\begin{aligned}B_{y}= & B_{0}\tanh\bigg\{\frac{1}{\delta}\bigg[\alpha+\cos\bigg(\frac{2\pi(x+\beta_{0}ct)}{\lambda}\bigg)\bigg]\bigg\}\\
E_{z}= & \beta_{0}B_{0}\tanh\bigg\{\frac{1}{\delta}\bigg[\alpha+\cos\bigg(\frac{2\pi(x+\beta_{0}ct)}{\lambda}\bigg)\bigg]\bigg\}
\end{aligned}
\end{equation}
where $\beta_{0}$ is the velocity of the wind normalized by the speed
of light $c$, and $\lambda$ is the wavelength of the stripes in
the wind. The dimensionless parameters $\delta$ and $\alpha$ are
such that the half thickness of the current sheet is $\Delta\approx\lambda\delta/(2\pi)$
(actually $\Delta=(\lambda\delta)/(2\pi\sqrt{1-\alpha^{2}})$ as shown
later in this paragraph), and $B_{y}$ averaged over one wavelength
is $\langle B_{y}\rangle_{\lambda}=B_{0}[1-2(\arccos\alpha)/\pi]$.
We use the variable $\xi=\frac{1}{\delta}\big[\alpha+\cos\big(2\pi(x+\beta_{0}ct)/\lambda\big)\big]$
for the phase of the electromagnetic field. The location of current
sheets is determined by setting $\xi=0$, i.e. $x+\beta_{0}ct=(\lambda/(2\pi))\arccos[-\alpha]$.
And the location of the transitional field is determined by setting
$\xi=\pm1$, i.e. $x+\beta_{0}ct=(\lambda/2\pi)\arccos[-\alpha\pm\delta]$.
By subtracting the coordinates we get the half thickness of the current
sheet $\Delta=(\lambda\delta)/(2\pi\sqrt{1-\alpha^{2}})$. To make
sure $\Delta\ll\lambda$, we need to have $\delta/\sqrt{1-\alpha^{2}}\ll1$.
The background cold electron/positron plasma in the wind is uniform,
with constant density $n_{e,p}^{\mathrm{cold}}=n_{c0}/2$ and constant
temperature $kT_{e,p}^{\mathrm{cold}}=0.04m_{e}c^{2}$. The hot electron/positron
plasma balances the magnetic pressure and keeps the steady profile
of the electromagnetic field. Besides frame $S$ which is also the
downstream rest, there are several reference frames, including the
center-of-mass (CM) frame $S^{\prime}$ of the wind, the CM frame
$S_{e}^{\prime\prime}$ of the hot electrons, and the CM frame $S_{p}^{\prime\prime}$
of the hot positrons. The space-time coordinate transform between
$S$ and $S^{\prime}$ is\cite{Landau1963}

\begin{equation}
\begin{cases}
x & =\gamma_{0}(x^{\prime}-\beta_{0}ct^{\prime})\\
t & =\gamma_{0}(t^{\prime}-\beta_{0}\frac{x^{\prime}}{c})
\end{cases}\qquad\begin{cases}
x^{\prime} & =\gamma_{0}(x+\beta_{0}ct)\\
t^{\prime} & =\gamma_{0}(t+\beta_{0}\frac{x}{c})
\end{cases}
\end{equation}
where $\gamma_{0}=1/\sqrt{1-\beta_{0}^{2}}$ is the Lorentz factor
for the transformation between $S$ and $S^{\prime}$.

The electromagnetic field in frame $S^{\prime}$ can be derived by
the Lorentz transform of electromagnetic field\cite{Landau1963}
\begin{equation}
\begin{aligned}E_{x}^{\prime} & =E_{x}=0\\
E_{y}^{\prime} & =\gamma_{0}(E_{y}+\beta_{0}B_{z})=0\\
E_{z}^{\prime} & =\gamma_{0}(E_{z}-\beta_{0}B_{y})=0\\
B_{x}^{\prime} & =B_{x}=0\\
B_{y}^{\prime} & =\gamma_{0}(B_{y}-\beta_{0}E_{z})=\frac{B_{0}}{\gamma_{0}}\tanh\xi=\frac{B_{0}}{\gamma_{0}}\tanh\bigg\{\frac{1}{\delta}\bigg[\alpha+\cos\bigg(\frac{2\pi x^{\prime}}{\lambda\gamma_{0}}\bigg)\bigg]\bigg\}\\
B_{z}^{\prime} & =\gamma_{0}(B_{z}+\beta_{0}E_{y})=0
\end{aligned}
\end{equation}
In $S^{\prime}$ frame, the electric field is zero everywhere, the
half thickness of the current sheet is $\Delta^{\prime}=\gamma_{0}\Delta$,
and the wavelength is $\lambda^{\prime}=\gamma_{0}\lambda$.

We assume $S_{p}^{\prime\prime}$ is moving at speed $\beta_{h}\hat{z}$
with respect to $S^{\prime}$. The charge density-current tensor $(\rho_{hp}^{\prime\prime},0,0,0)$
in $S_{p}^{\prime\prime}$ frame can be transformed into the charge
density-current tensor $(\rho_{hp}^{\prime},j_{hp,x}^{\prime},j_{hp,y}^{\prime},j_{hp,z}^{\prime})$
in $S^{\prime}$ frame and $(\rho_{hp},j_{hp,x},j_{hp,y},j_{hp,z})$
in $S$ frame
\begin{equation}
\begin{cases}
\rho_{hp}^{\prime} & =\gamma_{h}(\rho_{hp}^{\prime\prime}+\frac{\beta_{h}}{c}j_{hp,z}^{\prime\prime})=\gamma_{h}\rho_{hp}^{\prime\prime}\\
j_{hp,x}^{\prime} & =j_{hp,x}^{\prime\prime}=0\\
j_{hp,y}^{\prime} & =j_{hp,y}^{\prime\prime}=0\\
j_{hp,z}^{\prime} & =\gamma_{h}(j_{hp,z}^{\prime\prime}+\beta_{h}c\rho_{hp}^{\prime\prime})=\gamma_{h}\beta_{h}c\rho_{hp}^{\prime\prime}
\end{cases}\qquad\begin{cases}
\rho_{hp} & =\gamma_{0}(\rho_{hp}^{\prime}+\frac{\beta_{0}}{c}j_{hp,x}^{\prime})=\gamma_{0}\gamma_{h}\rho_{hp}^{\prime\prime}\\
j_{hp,x} & =\gamma_{0}(j_{hp,x}^{\prime}+\beta_{0}c\rho_{hp}^{\prime})=\gamma_{0}\beta_{0}\gamma_{h}c\rho_{hp}^{\prime\prime}\\
j_{hp,y} & =j_{hp,y}^{\prime}=0\\
j_{hp,z} & =j_{hp,z}^{\prime}=\gamma_{h}\beta_{h}c\rho_{hp}^{\prime\prime}
\end{cases}
\end{equation}
where $\gamma_{h}=1/\sqrt{1-\beta_{h}^{2}}$. The velocity of $S_{p}^{\prime\prime}$
in $S$ is 
\begin{equation}
(v_{hp,x},v_{hp,y},v_{hp,z})=(j_{hp,x}/\rho_{hp},j_{hp,y}/\rho_{hp},j_{hp,z}/\rho_{hp})=c(\beta_{0},0,\beta_{h}/\gamma_{0})
\end{equation}
Assuming the density of the hot electrons/positrons in the current
sheet in $S$ frame is $n_{he,hp}=n_{h0}/(2\cosh^{2}\xi)$, where
$n_{h0}/n_{c0}=\eta$ is the overdensity relative to the cold particles
outside the layer and is set to be $\eta=3$\cite{Striped_Kirk2003,Sironi2011,Reconn_Sironi2014}.
Then in $S_{he,hp}^{\prime\prime}$ frame we have $n_{he,hp}^{\prime}=n_{h0}/(\gamma_{0}\gamma_{h})/(2\cosh^{2}\xi)$.

The temperature of hot component is $kT_{h}=\eta_{T}m_{e}c^{2}$.
In $S^{\prime}$ frame, the sum magnetic pressure and hot component
thermal pressure in $x$ direction is
\begin{equation}
p_{x}^{\prime}=\frac{B_{y}^{\prime2}}{8\pi}+n_{he}^{\prime\prime}k_{B}T_{h}+n_{pe}^{\prime\prime}k_{B}T_{h}=\frac{B_{0}^{2}}{8\pi\gamma_{0}^{2}}\tanh^{2}\xi+\frac{n_{c0}\eta\eta_{T}m_{e}c^{2}}{\gamma_{0}\gamma_{h}\cosh^{2}\xi}
\end{equation}
To ensure pressure balance we need $B_{0}^{2}/(8\pi\gamma_{0}^{2})=(n_{c0}\eta_{n}\eta_{T}m_{e}c^{2})/(\gamma_{0}\gamma_{h})$,
thus $\eta_{T}=\frac{\sigma_{0}\gamma_{h}}{2\eta_{n}}$ where the
magnetization parameter is $\sigma_{0}=\frac{B_{0}^{2}}{4\pi\gamma_{0}n_{c0}m_{e}c^{2}}$.

The drift velocity $\beta_{h}\hat{z}$ of the hot particles is determined
by keeping the steady profile of electromagnetic field, i.e. in the
rest frame of the wind $\nabla\times\boldsymbol{B}^{\prime}=(4\pi/c)\boldsymbol{J}^{\prime}$
is satisfied so that the electric field stays zero. Thus 
\begin{align}
J_{z}^{\prime} & =-\rho_{he}^{\prime\prime}c\beta_{h}\gamma_{h}+\rho_{hp}^{\prime\prime}c\beta_{h}\gamma_{h}e=\frac{\eta n_{c0}c\beta_{h}\cosh^{-2}\xi}{\gamma_{0}}\nonumber \\
 & =\frac{1}{4\pi}\partial_{x^{\prime}}B_{y}^{\prime}=-\frac{B_{0}}{2\lambda\gamma_{0}^{2}\delta}\cosh^{-2}\xi\sin\bigg(\frac{2\pi x^{\prime}}{\lambda\gamma_{0}}\bigg)
\end{align}
Thus
\begin{equation}
\beta_{h}=-\frac{B_{0}}{2\lambda\gamma_{0}\delta e\eta_{n}n_{c0}c}\sin\bigg(\frac{2\pi(x+\beta_{0}ct)}{\lambda}\bigg)=-\frac{\sin\bigg(\frac{2\pi(x+\beta_{0}ct)}{\lambda}\bigg)}{\sqrt{1-\alpha^{2}}}\frac{\sqrt{\sigma}(c/\omega_{p})}{(\eta_{n}\gamma_{0})\Delta}
\end{equation}
where $\omega_{p}=\sqrt{4\pi n_{c0}e^{2}/(\gamma_{0}m_{e})}$ is the
plasma frequency of the cold background plasma. The time in our simulation
is normalized by $1/\omega_{p}$, and the spatial coordinates in our
simulation are normalized by $c/\omega_{p}$.

The boundary at $x=0$ is reflecting for particles and conducting
for fields. The shock is self-consistently generated by the interaction
between the reflected flow and the incoming flow. The simulation is
periodic in $y$ direction. In the run we show in this paper, we have
$\alpha=0.1$, $\Delta=d_{e}$, $\lambda=640d_{e}$, $\gamma_{0}=10^{4}$
and $\sigma_{0}=10$. The length of the simulation box in $y$ direction
is $L_{y}=400d_{e}$.

\section{Results}

\begin{figure}
\includegraphics[scale=0.6]{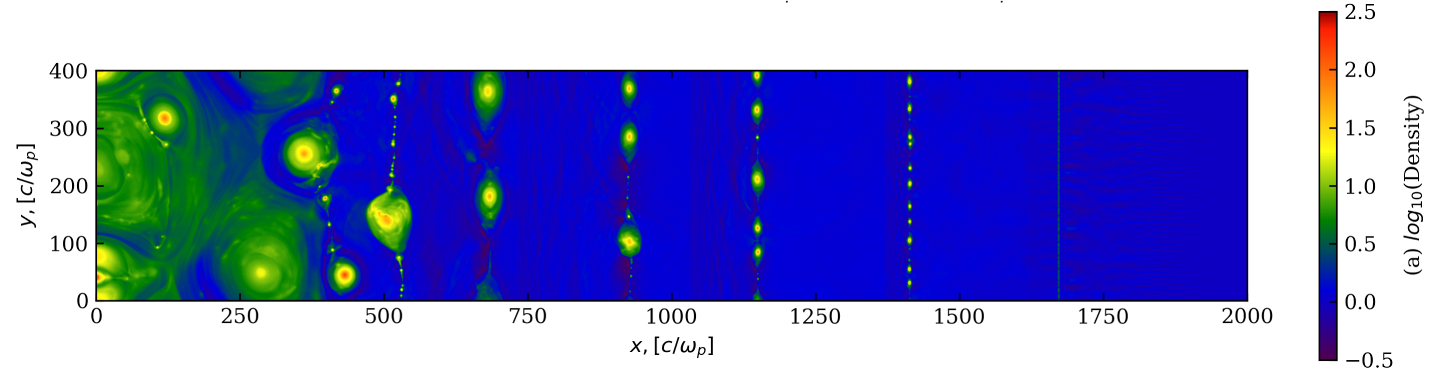}

\includegraphics[scale=0.5]{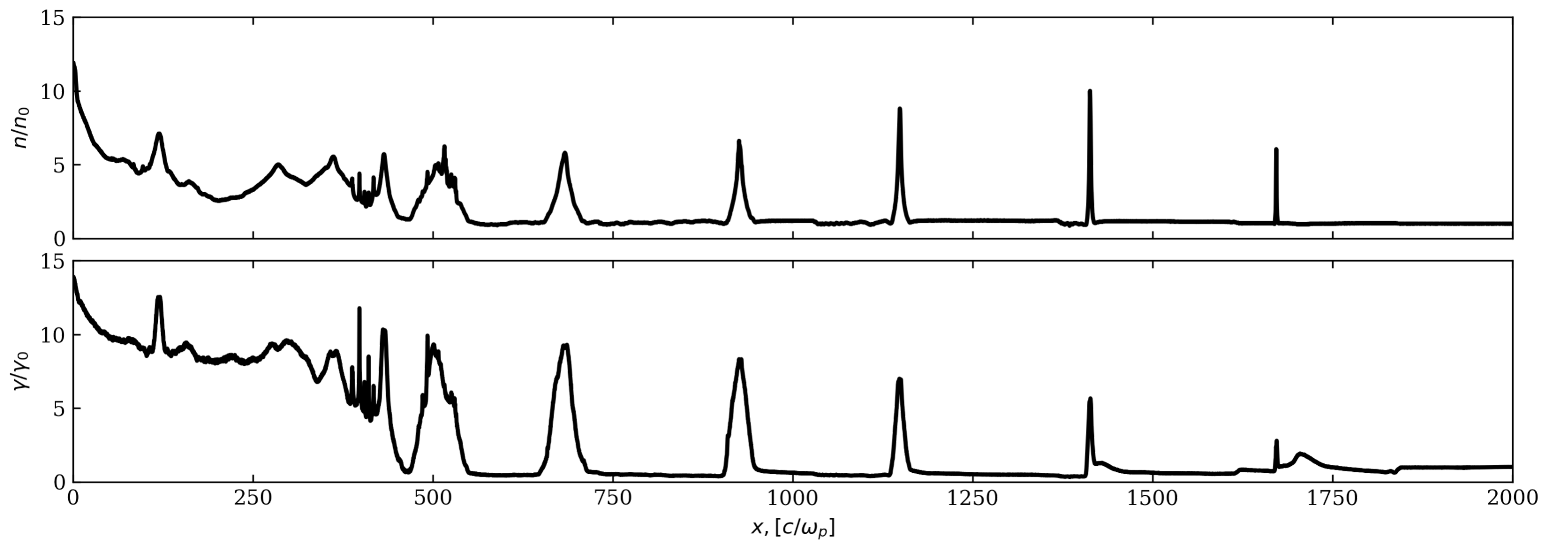}

\caption{Top: 2D spatial profile of logarithm of particle number density normalized
by $n_{c0}$ at time $\omega_{p}t=2000$. Middle: 1D spatial profile
(averaged over $y$) of particle number density normalized by $n_{c0}$
at time $\omega_{p}t=2000$. Bottom: 1D spatial profile of average
$\gamma$ of all particles normalized by $\gamma_{0}$ at time $\omega_{p}t=2000$.
\label{fig:spatial}}
\end{figure}

\begin{figure}
\includegraphics[scale=0.4]{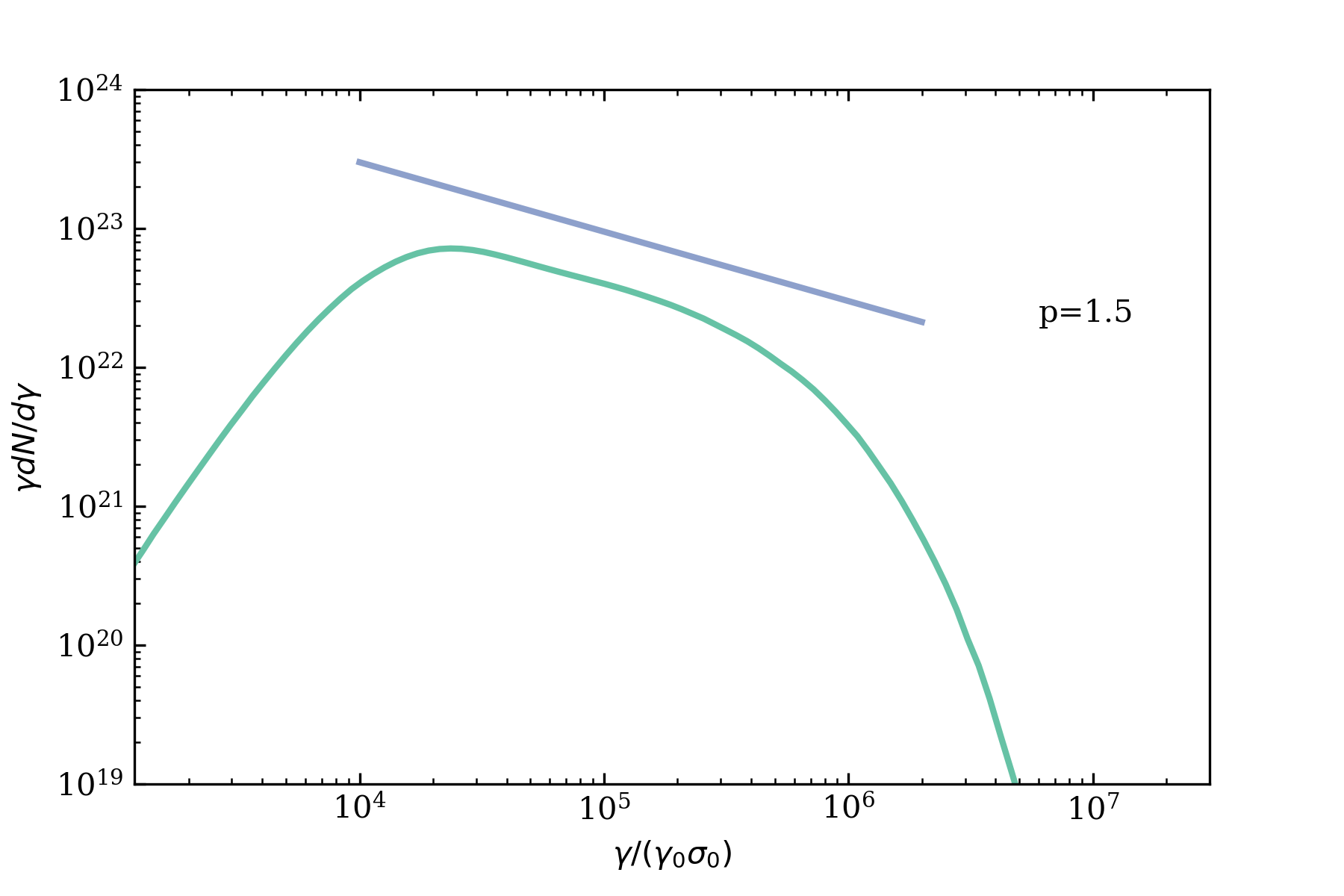}\includegraphics[scale=0.4]{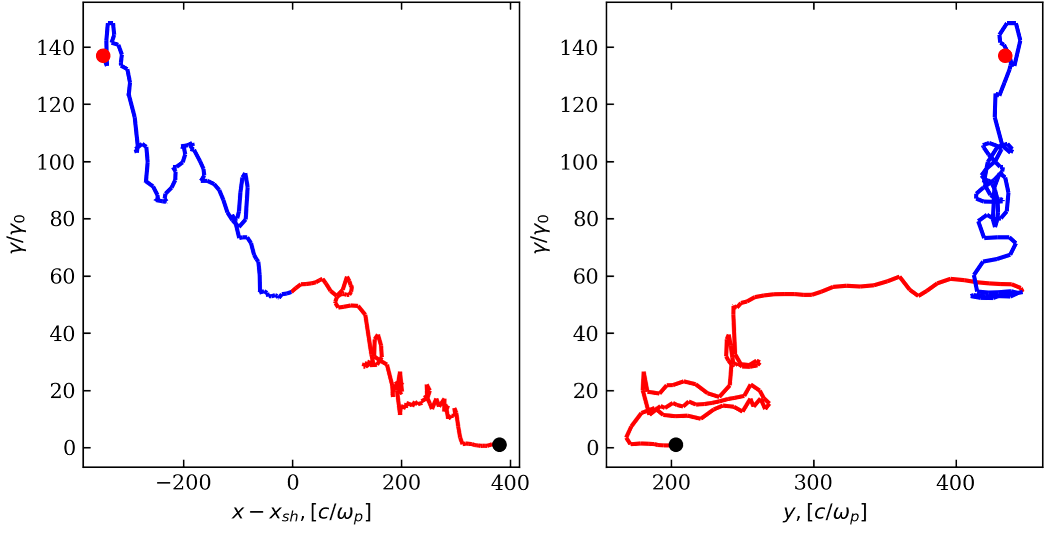}

\caption{Left: particle energy distribution $\gamma dN/d\gamma$ for all the
electrons and positrons in downstream region $0<x<354c/\omega_{p}$
at $\omega_{p}t=2000$. Middle and right: particle trajectory for
a particle in the simulation. The middle column is particle Lorentz
factor (normalized by $\gamma_{0}$) vs. $x-x_{\mathrm{sh}}$ coordinate
where the location of the shock jump is approximately at $x_{\mathrm{sh}}\approx(ct-1000c/\omega_{p})/(2\sqrt{2})$,
and the right column is particle Lorentz factor (normalized by $\gamma_{0}$)
vs. $y$ coordinate. The red line is the trajectory where the particle
is in the upstream, and the red line is the trajectory where the particle
is in the downstream. The black dots are for particles at $t=0$,
and the red dots are for particles at $\omega_{p}t=2000$. \label{fig:particle}}
\end{figure}

As shown in Figure \ref{fig:spatial}, the shock forms self-consistently,
and propagates to the right side of the simulation box, compresses
and decelerates the current sheets. The results are consistent with
the previous study\cite{Sironi2011}. The shock converts the magnetic
energy into the particle kinetic energy, resulting in an average particle
Lorentz factor $\langle\gamma\rangle\approx\gamma_{0}(1+\sigma_{0})$
in the downstream, which is consistent with the jump condition of
ultra-relativistic magnetized shock. The location of the shock is
approximately $x_{sh}\approx(ct-1000c/\omega_{p})/(2\sqrt{2})$, moving
at speed approximately $c/(2\sqrt{2})$. In the upstream, the current
sheets start to continuously break into a series of magnetic islands
separated by X-points. The islands coalesce, grow to larger size,
and further grow after passing the shock front.

The energy distribution function at downstream of the shock follows
a power law $f(\varepsilon)\propto\varepsilon^{-p+1}$ with $p=1.5$
as shown in Figure \ref{fig:particle}. The trajectory for a typical
tracer particle that feels the self-consistent electromagnetic field
is shown in Figure \ref{fig:particle}. The trajectories of most high-energy
tracer particles are similar to the one in Figure \ref{fig:particle}.
Those tracer particles (i) bounce several times in the upstream and
then travel into the downstream, or (ii) travel into the downstream
and bounce several times in the downstream, or (iii) do (i) followed
by (ii). Few particles get bounced between upstream and downstream,
i.e after a particle travels into the downstream of the shock it rarely
travels back into the upstream. The fact that we get a $p=1.5$ power
law is consistent with the picture that Fermi mechanism in the reconnection
islands is much more efficient than the diffusive shock acceleration.

\section{Conclusions and discussions}

While a growing body of research using PIC methods focus on the particle
acceleration in the spontaneous relativistic magnetic reconnection
in the magnetically dominated regime, in this work we extended the
study by setting up a simulation with shock driven magnetic reconnection
at the termination shock of relativistic striped wind. The analysis
shows that many particles are accelerated by Fermi-type mechanism.
It is known that PIC simulations only covers a small region in the
termination shock, while in reality there is enormous scale separation
between the system size and the dynamical scale such as $c/\omega_{p}$.
However, to model the particle acceleration during magnetic reconnection
in a macroscopic system, it is important to determine the dominant
acceleration mechanism.

\begin{acknowledgement}

Research presented in this paper was supported by the Center for Space
and Earth Science (CSES) program, Laboratory Directed Research and
Development (LDRD) program 20200367ER of Los Alamos National Laboratory
(LANL) and NASA ATP program through grant NNH17AE68I. The research
by P. K. was also supported by CSES. CSES is funded by LANL's LDRD
program under project number 20180475DR. The simulations were performed
with LANL Institutional Computing which is supported by the U.S. Department
of Energy National Nuclear Security Administration under Contract
No. 89233218CNA000001.

\end{acknowledgement}

\bibliography{/Users/yclu/GoogleDrive/Desktop/PWNE/LANL/PWNE}

\end{document}